\documentclass[twocolumn]{article}%
\usepackage{amsmath}
\usepackage{subcaption}
\usepackage{graphicx}
\graphicspath{{figures/}} 
\usepackage{esvect}
\usepackage{lipsum}
\usepackage[english]{babel}
\usepackage{braket}
\usepackage{verbatim}
\usepackage{amssymb}%
\usepackage[utf8]{inputenc}
\usepackage{enumitem}
\usepackage[font=small,labelfont=bf]{caption}
\usepackage[section]{placeins} 
\usepackage[
	left=2.00cm,
	right=2.00cm,
	top=2.00cm,
	bottom=2.00cm
]{geometry}
\usepackage{booktabs} 
\usepackage{siunitx}
\DeclareSIUnit\inch{in}
\DeclareSIUnit\photon{photons}
\DeclareSIUnit\micron{\mu m}
\usepackage{tabularx}

\usepackage[
	backend=biber,
	style=phys,
	sorting=none,
	maxnames=1
]{biblatex}
\renewbibmacro{in:}{} 
\usepackage[colorlinks,citecolor=red,urlcolor=blue,bookmarks=false,hypertexnames=true]{hyperref} 
\usepackage[capitalise,noabbrev]{cleveref} 
\AtEveryBibitem{
	\iffieldequalstr{collaboration}{DUNE}{\printnames{author}\space\printtext{(}\printfield{collaboration}\printtext{)}\newunitpunct\printfield{title}\addcomma\clearname{author}}{} 
	\clearfield{title} 
}

\usepackage{lineno} 
\newcommand{\order}[1]{\mathcal{O}(#1)}

\begin{document}

\title{Measurement of the Angular Distribution of Wavelength-Shifted Light Emitted by TPB}
\author{J. Schrott$^{1,2}$, M. Sakai$^{1,2}$, S. Naugle$^{1,2}$, G. D. Orebi Gann$^{1,2}$,\\
S. Kravitz$^{1,2}$, D. McKinsey$^{1,2}$, R.J. Smith$^{1,2}$}
\date{$^1${\footnotesize University of California, Berkeley} \\
$^2${\footnotesize Lawrence Berkeley National Laboratory}\\[2ex]
July 2020}
\maketitle

\begin{abstract}
We present measurements of the angular distribution of re-emitted light from tetraphenyl
butadiene thin films when exposed to \SI{128}{nm} light, which is the peak of the liquid Argon (LAr) scintillation spectrum, in vacuum. Films
ranging from \SI{250}{nm} to \SI{5.5}{\micron} in thickness are measured.
All films were fabricated by evaporation deposition on ultraviolet transmitting (UVT) acrylic substrates.
Preliminary comparisons of the angular distribution to that produced by a detailed Monte Carlo model are also presented. The current
shortcomings of the model are discussed and future plans briefly outlined.
\end{abstract}

\maketitle

\section{Introduction}
\label{sec:introduction}

Noble liquid (NL) radiation detectors are a widely used tool in modern particle physics experiment. These detectors enable a variety of
research programs ranging from neutrino
physics~\cite{Abi:2020kei,Acciarri:2016smi}
and dark matter
searches~\cite{Bernstein_2020,Tvrznikova:2019tgx,Katsioulas:2018squ,Baudis:2014naa,Rielage:2014pfm,Schumann:2012ph,Manalaysay:2011ix,Agnes:2015ftt,Abe:2013tc,Baudis:2012bc,Akerib:2019fml,Akerib:2016vxi,Aprile:2019xxb,Calvo:2016hve,Brunetti:2004cf,Boulay:2012hq,Akimov:2006qw,Aprile_2012}
to other rare-event measurements~\cite{Wittweg:2020fak,Anton:2019wmi,Agostini:2020adk,Raj:2019sci,Bondar:2017til,Baxter:2017ozv,Chepel:2012sj,Albert:2017owj,Shakeri:2020wvk,Amerio:2004ze,Myslik:2018knv}.
In recent years, significant effort has been committed to exploring how
large-scale NL detectors can answer pressing questions in these fields.
Numerous such detectors have been proposed, and several are already operational
or projected to be in the near future%
~\cite{
  Abi:2020kei,
  Akerib:2019fml,
  Aprile_2012,
  Myslik:2018knv,
  Baxter:2017ozv,Kravitz:2019zqv}. 
Unsegmented monolithic NL detectors enjoy better self shielding than water or organic scintillators and have typical photon yields of
\SIrange[range-units=single]{20000}{40000}{\photon\per\mega\electronvolt},
leading to excellent energy resolution and low energy thresholds. Because of the particular process by which NLs scintillate,
involving excited dimer molecules, they are also highly transparent to their own scintillation light and therefore make very scalable detectors. Yet another feature are the long recombination times observed of free electrons produced when charged particles collide with and ionize atoms in an NL. These electrons can drift large distances through a liquid volume when an electric field is applied, allowing NL detectors to be used as so-called time projection chambers~\cite{Abi:2020kei}.

In concert with the research efforts mentioned above, there are pushes to better
understand materials that comprise the inner volumes of NL detectors. As
designs are refined, accurate modeling of the optics of these materials becomes
an important limiting factor in searching for new physics. In particular, those
materials which strongly affect light collection must be well-understood.

\begin{figure}[t!]
	\centering
	\includegraphics[width=\linewidth, keepaspectratio]{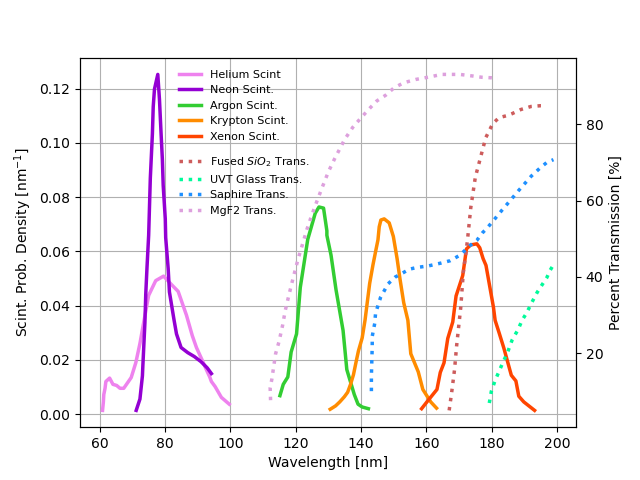}
	\caption{
    Scintillation spectra for various noble gases. The spectral transmittance
    of materials used in some common optical windows are also
    shown~\cite{Gehman:2011xm}.
	}
	\label{fig:lng_emission}
\end{figure}

\begin{figure} [t!]
	\centering
	\includegraphics[width=\linewidth, keepaspectratio]{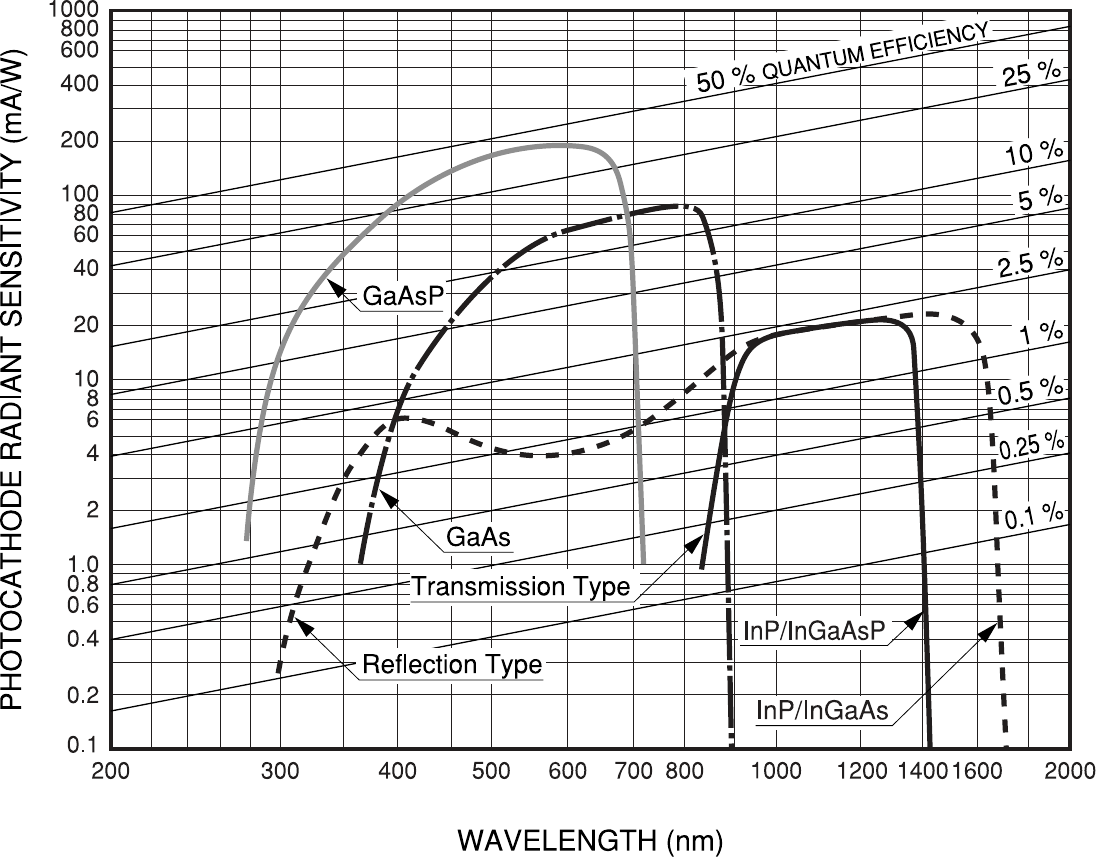}
	\caption{
    Typical radiant sensitivities of some common photo-cathode
    materials~\cite{Hamamatsu:website}.
	}
	\label{fig:photocathodes}
\end{figure}

Despite their wide use, NL detectors are burdened by the fact their scintillation light is deep in the ultraviolet (UV)
wavelength regime (\SI{178}{nm} for Xenon and down to approximately
\SI{80}{nm} for Helium or Neon). \cref{fig:lng_emission} shows the
scintillation emission spectra of several NLs as well as the transmittances of
some materials commonly used in optical windows ~\cite{Benson:2017vbw}.
Light below ~$\SI{160}{nm}$ is heavily absorbed by most optical windows and
cannot be detected by typical photo-multiplier tubes
(PMTs).
\cref{fig:photocathodes} shows the wavelength dependent sensitivities of some
common photo-cathode materials~\cite{Hamamatsu:website}.
A critical element of some NL detectors is the use of wavelength-shifting materials
to absorb and re-emit
scintillation light at more readily detectable wavelengths.

\begin{figure} [t]
	\centering
	\includegraphics[width=\linewidth, keepaspectratio]{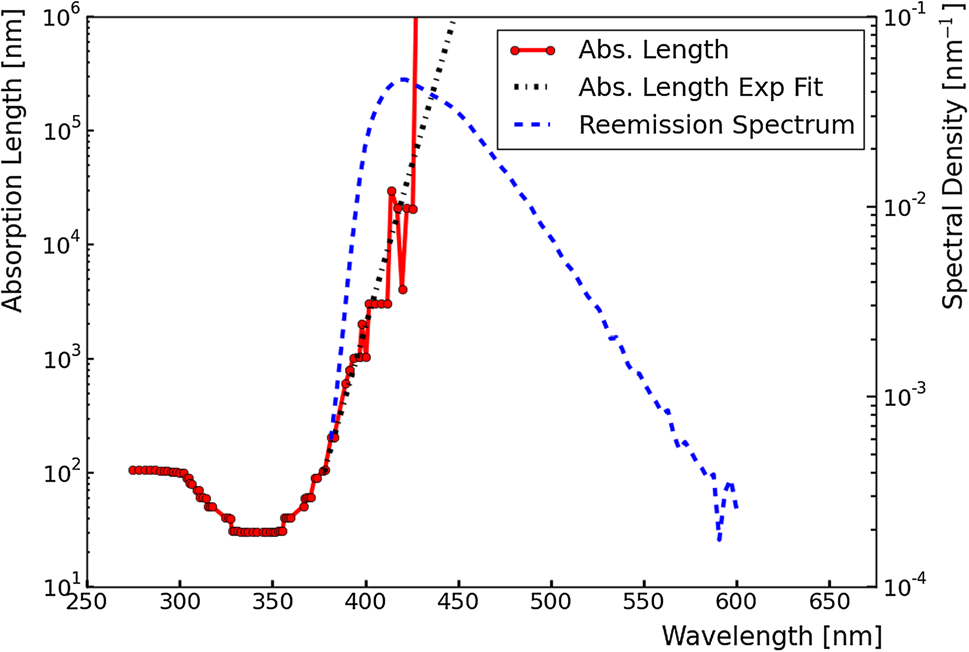}
	\caption{
		Spectral absorption length (red) and re-emission spectral density (blue) of
		TPB~\cite{Benson:2017vbw}.
	}
	\label{fig:reemission_absorption}
\end{figure}

A popular wavelength shifter is tetraphenyl butadiene (TPB), which can be deployed in numerous ways in the inner volume of an NL detector. Incoming scintillation photons
are absorbed by TPB and then re-emitted at detectable wavelengths
before they are incident on a PMT. \cref{fig:reemission_absorption} shows both
the re-emission spectrum and wavelength dependent absorption length of
TPB~\cite{Benson:2017vbw}. 

Previous studies have produced measurements of various optical properties of
TPB including the absorption and emission spectra, timing of re-emitted light,
and intrinsic quantum efficiency of a single TPB molecule
~\cite{Gehman:2013kra,Yang:2019zob,Benson:2017vbw,Gehman:2011xm,samson1967vacuum,Burton:1973tla,wallace1994excited,mckinsey1997fluorescence,Jones:2012hm,Francini:2013lua,Stolp:2016rde,regan1994measured,allison1964absolute,bruner1969absolute,Segreto:2015timing}.
However, the angular distribution of re-emitted light has not been
fully explored. This paper presents the first
measurement of the full angular distribution of re-emitted light from
evaporation-deposited TPB films on acrylic when exposed to \SI{128}{nm} light, which is the peak of the liquid argon (LAr) scintillation spectrum. Past measurements
of the relative intensity of transmitted and reflected light through a TPB
film on a silicon substrate have been made~\cite{Stolp:2016rde}, and these measurements are consistent with the data taken in this study. In this paper,
measurements of the angular distribution of re-emitted light are made for multiple TPB film thicknesses ranging from \SI{250}{nm} to \SI{5.5}{\micron}. 

Both the integral of the angular distribution of re-emitted light and the ratio of transmitted and reflected light from a TPB film as functions of film thickness are of interest for those hoping to maximize the light
collection efficiencies of their detectors. In a detector configuration
with TPB applied to the outer surface of a PMT window, the ideal behavior of
the TPB is to maximize the (shifted) transmitted light. In addition to being a useful measurement in its own right, the angular
distribution of re-emitted light can in theory be used to extract
important micro-physical optical properties of TPB via comparison to a
detailed Monte Carlo (MC) model. Therefore these measurements can inform how TPB is modeled in future optical simulations. The methodology of such a MC study is
outlined in \cref{sec:monte_carlo}.

\cref{sec:apparatus} provides an overview of the hardware and basic operation
of the experimental apparatus. \cref{sec:sample} details the fabrication of TPB
film samples. \cref{sec:measurements} discusses the measurements used to
account for backgrounds and degradation of the optical components from UV exposure. \cref{sec:results} presents the data and results from this study. \cref{sec:monte_carlo} presents preliminary comparisons of the data to the angular distributions produced by a micro-physical MC model. \cref{sec:conclusions,sec:acknowledgements} are conclusions and acknowledgments respectively.

\section{Experimental Apparatus}
\label{sec:apparatus}

\begin{figure} [t]
	\centering
	\includegraphics[width=\linewidth, keepaspectratio]{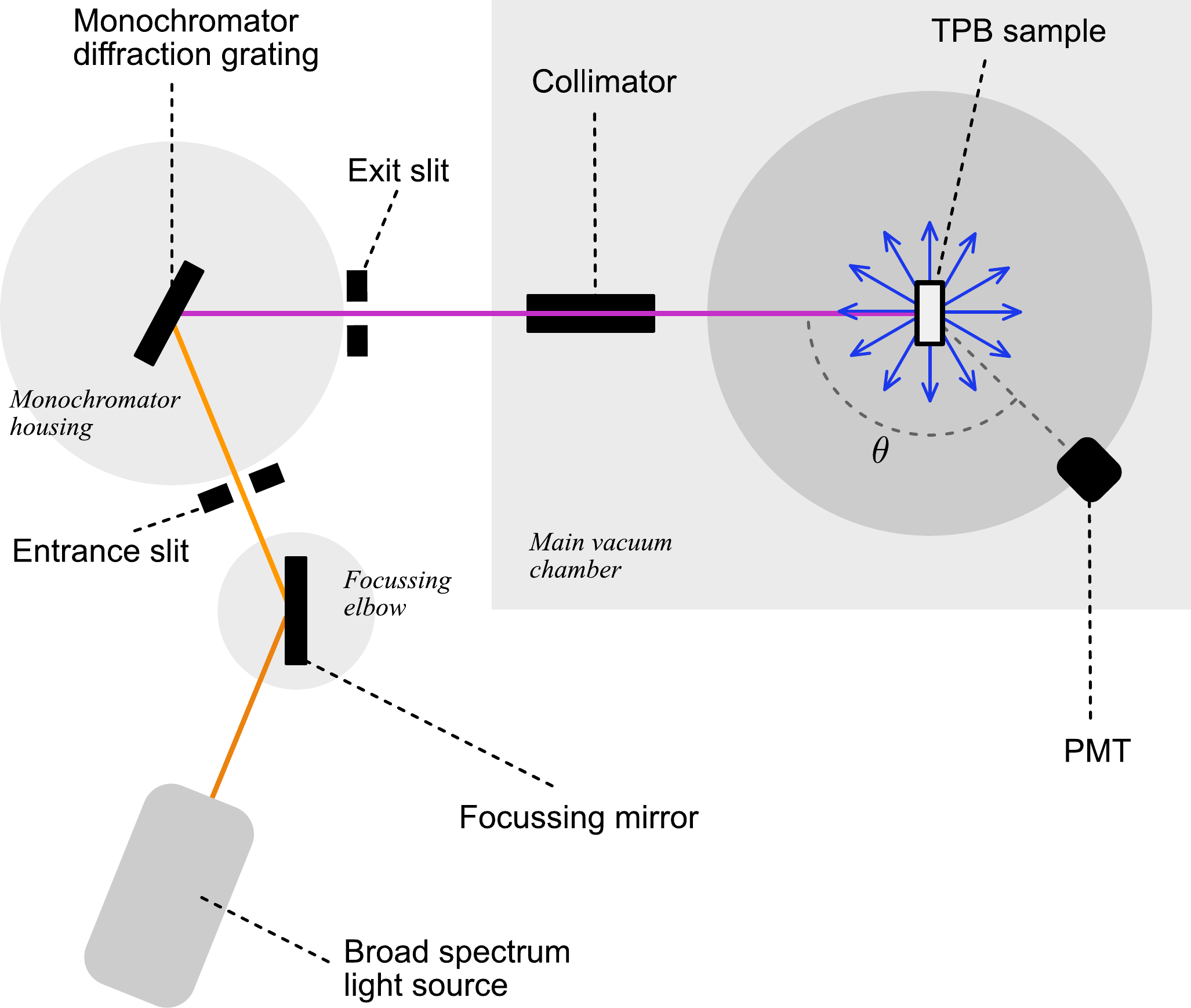}
	\caption{
		Schematic of the experimental apparatus when viewed from above. The
		orange line represents the beam of broad spectrum light
		produced at the source and focused onto the diffraction grating. The purple line represents the
		monochromatic, 128nm beam illuminating the TPB film. The blue arrows diverging from
		the sample represent the wavelength shifted light after absorption and
		re-emission by the TPB. The PMT viewing angle $\theta$ can be adjusted with a stepper motor.
	}
	\label{fig:schematic}
\end{figure}

This section describes the optical elements used to produce the monochromatic
light incident on the TPB samples. The apparatus used in this study builds upon
two former experiments conducted at Lawrence Berkeley National Laboratory;
namely a previous study of the micro-physical optical properties of
TPB~\cite{Benson:2017vbw} and a more recent project known as the Immersed
BRIDF (bi-directional reflectance intensity distribution function) Experiment in Xenon (IBEX), which investigated the optical reflectivity of
PTFE in both vacuum and liquid xenon~\cite{Kravitz:2019zqv}. The apparatus used
in this study is the same as that used in the
latter, and the basic arrangement of the optical chain remains unchanged.
\cref{fig:schematic} shows a schematic of the apparatus. The optical chain
begins with a McPherson Model 632 light source~\cite{McPherson:632}, which
produces light in the \SIrange{115}{400}{nm} wavelength range. This broad
spectrum light is focused with a McPherson Model 615 focusing elbow through a
vertical slit and onto a McPherson Model 234/302 Vacuum Ultraviolet
Monochromator, which can be adjusted to reflect monochromatic light of a
desired wavelength into the main vacuum chamber. The wavelength distribution of the resulting beam is less than $\SI{10}{nm}$ full width at half maximum.
In the main vacuum chamber,
the monochromatic light is collimated before hitting
the TPB sample housed in an aluminum sample holder. Pictures of the apparatus are provided in Kravitz (2020)~\cite{Kravitz:2019zqv}. The entire optical chain is
housed in vacuum.

\begin{figure} [t]
	\centering
	\includegraphics[width=\linewidth, keepaspectratio]{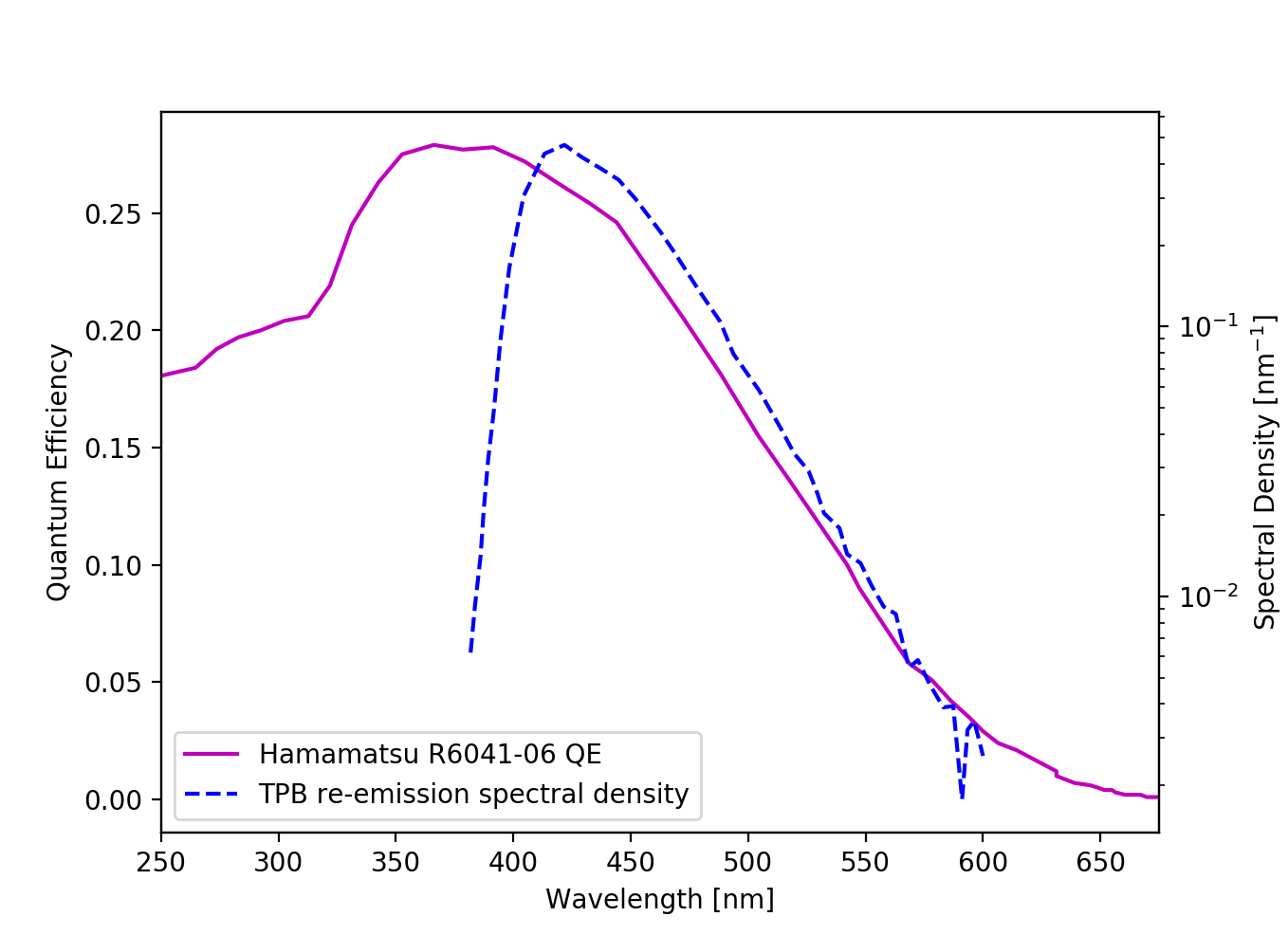}
	\caption{
		Quantum efficiency of the Hamamatsu R6041-06 photo
		multiplier tube (purple) along with the re-emission spectral density of TPB (blue)~\cite{Hamamatsu:website}.
	}
	\label{fig:pmt_efficiency}
\end{figure}

In the TPB film, the UV light may be absorbed and re-emitted. The re-emitted
light is wavelength shifted and distributed according to the spectrum shown in
\cref{fig:reemission_absorption}. This re-emitted light can then be collected by a
Hamamatsu R6041-06 PMT whose viewing angle relative to the beam path can be
varied from \SIrange{25}{185}{\degree}, with \SI{180}{\degree} corresponding to
the PMT facing straight into the beam if the TPB sample were removed from the optical chain. The aperture of the PMT is $\SI{0.965}{cm}$ in diameter and subtends a solid angle of $4\pi\cdot3\times10^{-4}$ from the center of the TPB film. The quantum efficiency of the PMT is shown in
\cref{fig:pmt_efficiency}. The adjustable PMT viewing angle
allows for the measurement of the angular distribution of re-emitted photons.

The TPB sample can also be rotated about the axis coming out of the page in
\cref{fig:schematic}.
While TPB films were exclusively positioned normal to the incident beam in this study,
the ability to rotate the sample holder is crucial for calibration of the sample
holder dial and validation of the MC geometry. The sample holder can
simultaneously hold four different samples aligned vertically. It can be
moved up and down to choose which sample to illuminate or to lift the entire holder out of the optical chain (all while under vacuum). 

Initial amplification of the raw PMT signal is done with a Phillips
Scientific Model 777 low noise amplifier. A LeCroy Model 623B discriminator
defines the triggering threshold and outputs square wave trigger pulses. The
trigger pulses are amplified once more and sent to a rate meter that integrates
and outputs an analog voltage logarithmically proportional to the rate of the
discriminator output. The signal from the rate meter output is finally
digitized and piped to a LabView module. Small corrections are made to
the signal in LabView to account for non-linearities between the PMT rate and
signal response up to $\order{\si{MHz}}$, which is roughly the maximum observed
rate in this apparatus. The threshold of the discriminator was set to a constant
value which attempts to minimize the number of triggers on false events and maximize
counting of real photon pulses.

Of all the elements in the optical chain, only the PMT rotation axis and sample holder rotation axis are fixed in place. The beam therefore had to be aligned so as to be incident on the center of the sample holder. A discussion of the alignment procedures used in this study is found in Kravitz (2020)~\cite{Kravitz:2019zqv}.

\section{Sample Fabrication}
\label{sec:sample}

TPB is an opaque organic material that forms small crystal domains in the solid
state. High purity TPB can readily be purchased in powder form. In this study,
films were fabricated on 1-inch diameter acrylic substrates by heated
evaporation deposition of pure TPB powder. The thickness of the films can be
controlled by varying the amount of evaporated TPB as well as the distance
between the substrate and heated TPB source. In this study, a Tectra Mini
Coater was used at a source-substrate distance of \SI{4.2}{\inch}.

\begin{figure} [t]
	\centering
	\includegraphics[width=1.1\linewidth, keepaspectratio]{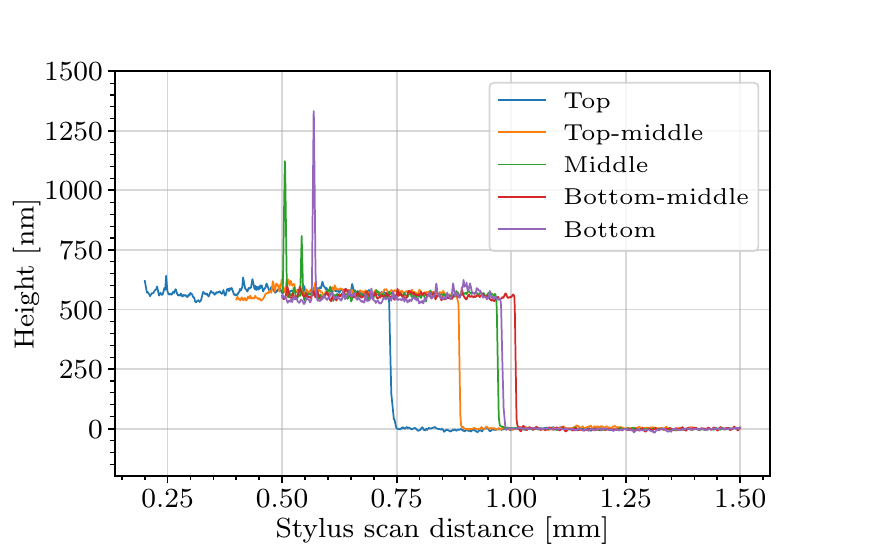}
	\caption{
		Thickness uniformity check of the TPB film. The thickness was
		measured at five locations on the same TPB film spaced in uniform
		intervals from one edge of the film (Top) through the middle of the
		film (Middle) to the diametrically opposing edge (Bottom). The
		measurements are overlaid here with arbitrary offsets along the
		horizontal axis. All five thickness measurements depicted by the height
		of the step are consistent with each other. The large spikes are likely caused by specs of dust on the TPB surface.
	}
	\label{fig:tpb_thickness_uniformity}
\end{figure}

To verify the scale, uniformity, and reproducibility of sample film
thicknesses, samples were scanned with a KLA-Tencor D-600 profilometer (also
known as Alpha-Step) at the Marvell Nanofabrication Laboratory at UC Berkeley. The profilometer uses a stylus that is
physically placed on the TPB surface and dragged across it to measure its
topography. Although the surface of an acrylic substrate can vary by
as much as \SI{100}{nm} from flat, it was found that evaporation deposition
nonetheless produces films of uniform thickness. This was confirmed by
fabricating a test sample with a thin strip of Kapton
tape covering the middle of the substrate. The TPB was deposited and the Kapton
tape removed to reveal a TPB-acrylic boundary that ran through the middle of
the sample. A profilometer step-height scan was made at five locations
uniformly spaced along this boundary, as shown in
\cref{fig:tpb_thickness_uniformity}, to confirm that the film thickness near the
center of the sample was consistent with that near the circumference. This
particular test showed a thickness of \SI{566 \pm 6}{nm} or, equivalently,
a variation of \SI{1}{\percent}.


The TPB films used for production data were coated on identical
1-inch diameter, circular UVT acrylic substrates in a 0.9-inch diameter area
concentric with the substrate. No Kapton tape was used for the fabrication of
the production data samples. The slightly smaller radius of the film leaves
the peripheral edge of the substrate exposed so profilometer
scans can be made at the TPB-acrylic boundary. These profilometer scans
were performed twice for each sample at
two diametrically opposite locations at the circumference of the film. The results of these two measurements were averaged to designate a single thickness for
each sample.
A list of TPB samples used in this study is presented
in~\cref{tab:list_of_tpb_samples} along with their measured thicknesses.


\begin{table} [t]
	\centering
	\caption[]%
	{%
		List of all TPB thin film samples and their measured thicknesses using
		a profilometer.
	}%
	\renewcommand{\arraystretch}{1.4}
	\begin{tabularx}{\linewidth}{@{}lcc@{}}
		\toprule%
		\textbf{Sample} & \textbf{Thickness (\si{\micron})} & \textbf{Uncertainty (\si{\micron})} \\
		\midrule%
		SN114 & \num{0.24} & $\substack{+0.07\\-0.07}$ \\ 
		SN105 & \num{0.57} & $\substack{+0.21\\-0.14}$ \\
		SN104 & \num{0.60} & $\substack{+0.13\\-0.13}$ \\
		SN107 & \num{1.31} & $\substack{+0.05\\-0.05}$ \\
		SN108 & \num{2.55} & $\substack{+0.07\\-0.06}$ \\
		SN111 & \num{3.68} & $\substack{+0.09\\-0.08}$ \\
		SN112 & \num{5.32} & $\substack{+0.28\\-0.28}$ \\
		SN113 & \num{5.39} & $\substack{+0.15\\-0.15}$ \\
		\bottomrule%
	\end{tabularx}
	\label{tab:list_of_tpb_samples}
\end{table}

The stylus force used during the profilometer scans can be adjusted. It was confirmed that using a force
of $\SI{1}{\milli\gram}\times\SI{9.8}{\meter\per\square\second}$ or less is weak enough to prevent damage of the TPB film. This was done by
comparing the angular distribution of re-emitted light from a TPB sample before and after scans using different forces. The re-emission performance of the film was
unaffected to within uncertainties by the scan. All film thickness scans for
samples presented in this study were performed within this permissible force
range.

\section{Measurements}
\label{sec:measurements}

\begin{figure} [t!]
	\centering
	\includegraphics[width=\linewidth, keepaspectratio]{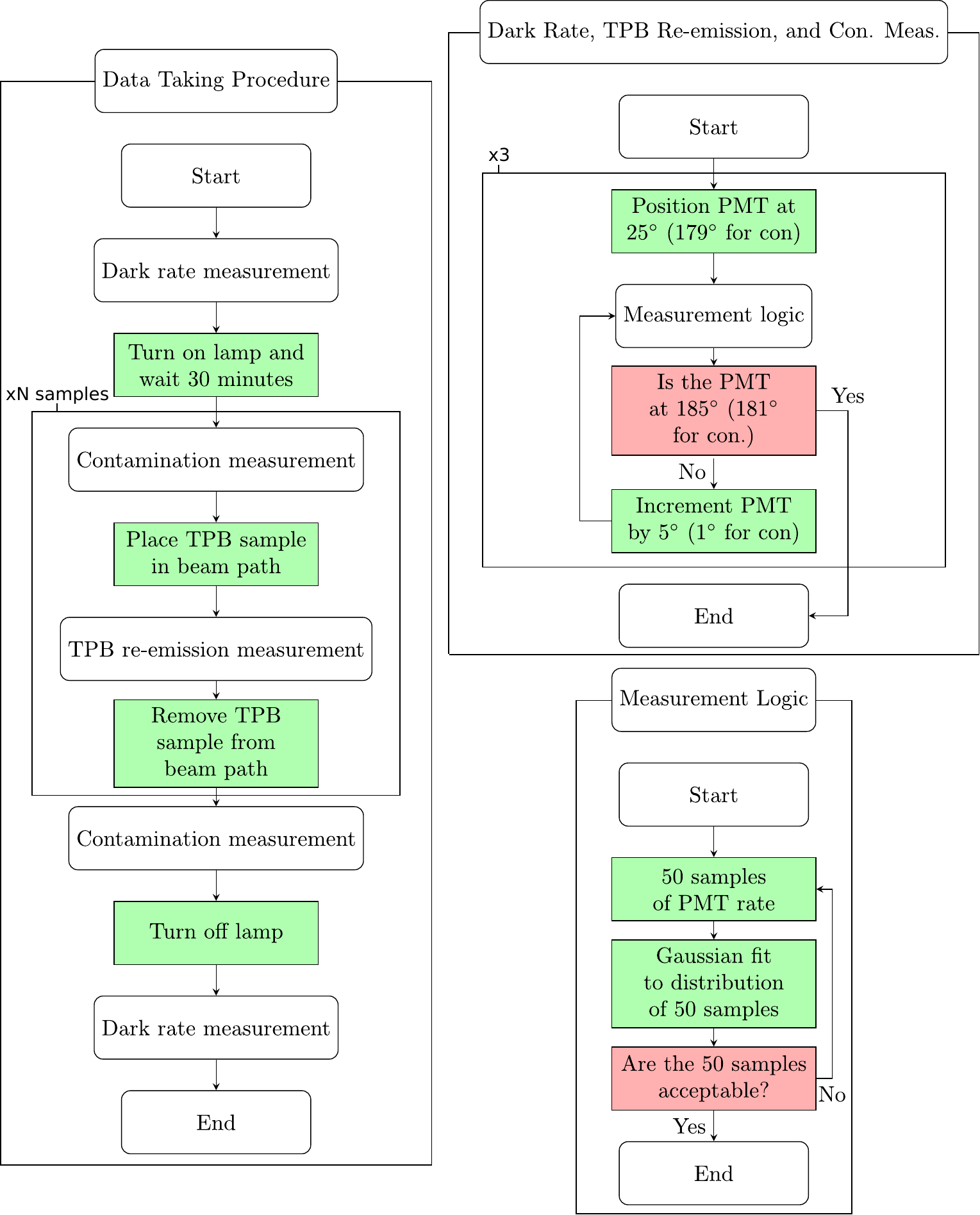}
	\caption{
	Flow chart summarizing the data taking procedure. 	}
	\label{fig:flowchart}
\end{figure}

A complete measurement of the angular distribution of re-emitted light from a TPB film involves three distinct types of measurement to account for backgrounds and systematic effects. 
A list of the different measurements is given below. A description of the full measurement procedure follows and a flowchart summarizing this procedure is given in \cref{fig:flowchart}.

\begin{itemize}
    \item \textit{\textbf{TPB Re-emission Measurement}}
    
    This is the signal measurement.  The PMT rate is measured every five degrees in viewing angle from \SIrange{25}{185}{\degree} 
    with the beam on and the TPB sample placed in front of the beam at normal incidence. 
    Typical rates at a viewing angle near \SI{180}{\degree} are $\order{\SI{10}{kHz}}$.
    
    \item \textit{\textbf{Dark Rate Measurement}}
    
    This is a PMT dark-rate and stray-light background measurement. 
    The PMT rate is measured every five degrees in viewing angle from \SIrange{25}{185}{\degree} 
    with the lamp off and the TPB sample lifted out of the main vacuum chamber.
Typical rates at any given viewing angle are $\order{\SI{10}{Hz}}$.
    
    \item \textit{\textbf{Contamination Measurement}}
    
    This measures the small amount of visible ``contamination" light in the UV beam and is used to monitor UV induced degradation of optical components (\cref{sec:lamp_deg,sec:tpb_deg}). 
The PMT rate is measured at \SI{179}{\degree}, \SI{180}{\degree}, and \SI{181}{\degree} with the beam on and the TPB sample lifted out of the main vacuum chamber. Typical rates at the viewing angle $\SI{180}{\degree}$ are $\order{\SI{1}{kHz}}$. 

\end{itemize}

\subsection{Data taking procedure}
\label{sec:data_taking}

Once the apparatus is under vacuum, the following steps are taken to realize a complete measurement of
the angular distribution of re-emitted light from a TPB film. 

First, a dark rate measurement is made at the start of any run to record the level of this background.  
After the dark rate measurement, the lamp is turned on and left to heat up for 30
minutes so its output is stable. With the lamp stabilized, a
contamination measurement is made.
After the contamination measurement, the desired TPB sample can be placed in
front of the beam at normal incidence. With the TPB in front of the beam, a TPB re-emission measurement is taken.
Immediately after the angular distribution measurement, the sample is removed from the beam path and another
contamination measurement is made, identical to the first. The average of the
integrals of the two contamination measurements is used to normalize TPB angular distributions taken at different times. Normalizing to contamination corrects for the degradation of the lamp window due to UV exposure (\cref{sec:lamp_deg}). 
Additional TPB re-emission measurements are performed, interleaved with contamination measurements, for a frequent monitor of potential UV degradation.  After data-taking is concluded, a second dark rate measurement is performed and checked for consistency with the first.

The following sections describe the background and systematic considerations in more detail.

\subsection{PMT stability}
\label{sec:pmt_stability}
The PMT exhibits instability on two time scales. The first and
simpler of the two is a slow, exponential decrease of the PMT dark rate from the
moment it is first turned on. This is dealt with by letting the PMT stabilize
for several hours before taking data. 

The second instability has a characteristic time on the order of tens of
seconds. It can be accounted for in the following way. For a single PMT rate
measurement, 50 PMT rate readings are sampled within
\(\lesssim\SI{1}{\second}\) and fit with a Gaussian in LabView. Ideally
these 50 measurements look Poissonian.  Occasionally, the distribution is
bi-modal, indicating that PMT instability has interfered with the measurement.
The criterion employed for the distribution having a well-defined central value is
that the mean of a fitted Gaussian must be within the interquartile rank of the
distribution.
If this is not the case, the 50 readings are re-sampled.  
As an additional cross check, for each scan across angles, the full scan is repeated three times and the results for a given angle averaged, with an additional uncertainty included to account for the spread between measurements.  This procedure is followed for signal, dark rate, and contamination measurements.



\subsection{Degradation of the deuterium lamp}
\label{sec:lamp_deg}
There are two time dependent degradation effects that affect rate measurements.
First is the degradation of the magnesium fluoride window of the deuterium lamp
due to polymerization of vacuum contaminants while the lamp is in operation.
The second is a gradual degradation of the TPB performance.
Even after significant use, the degradation of the deuterium lamp can be
reversed by following a window cleaning procedure available from its
manufacturer, McPherson. However, the rate of degradation during use was found
to be much greater just after a cleaning than after many hours of use. It was
crucial to ensure the lamp intensity remained stable over the course of a
single angular distribution measurement. Because the lamp output intensity was
adequate for measurements even after some degradation, it was decided
to let the lamp degrade to a point of relative stability where the change in
lamp output over the course of one angular distribution measurement was much
smaller than the Poisson statistics of the measurement. To confirm
the lamp output had advanced to a point of adequate stability, three
consecutive angular distribution measurements were compared and found to agree
within Poisson uncertainties.

Although this strategy ensures lamp degradation is negligible over the course
of one measurement of the angular distribution for one sample, after a
number of such measurements 
the effect can become important. A strategy was
therefore needed to normalize the overall integral of angular distributions
taken for different samples at different times to account for long term lamp degradation. 

A useful measurement for handling this is the photon rate seen by the PMT when
looking directly into the unobstructed \SI{128}{nm} beam, referred to in this study as a contamination measurement.
Although the PMT is not sensitive to the dominant \SI{128}{nm} light from the beam, it is sensitive to a small amount of longer wavelength ``contamination'' light which leaks in through the monochromator slit.

\begin{figure} [tbh!]
	\centering
	\includegraphics[width=1.1\linewidth, keepaspectratio]{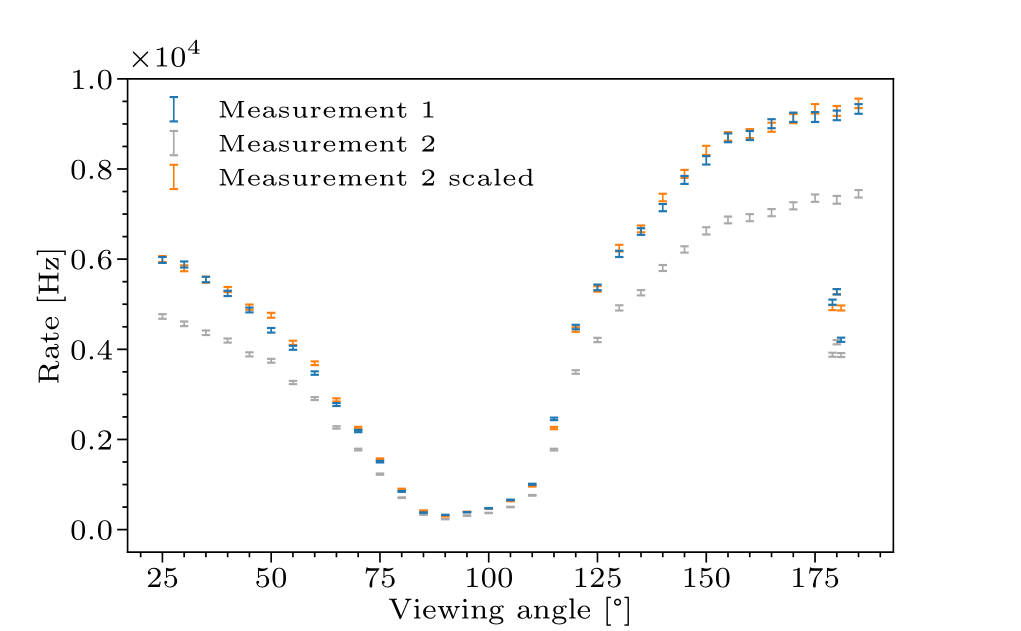}
	\caption{
		Angular distributions of re-emitted light from a TPB film taken before (blue) and after (gray) significant lamp window degradation. The cluster of points with lower intensity at viewing angles 179,180, and 181 are the contamination measurements taken at the same time as each angular distribution measurement. By scaling the ``after" measurement (gray) data such that the sum of the contamination measurements is the same for before and after, we see excellent agreement between the before and after data. This establishes that normalizing to contamination light is justified.
	}
	\label{fig:tpb_reemission_scaling_with_contam}
\end{figure}

It was observed that the intensity of this contamination light
co-varied with the intensity of the dominant \SI{128}{nm} light as a function
of lamp degradation. This was concluded by taking angular TPB re-emission
distribution measurements with the same sample before and after significant lamp
degradation and normalizing to the intensities of the contamination light
taken at the two times. When normalized to the
contamination light intensity, the angular distributions agree very well
as shown in~\cref{fig:tpb_reemission_scaling_with_contam}. Thus
contamination measurements of the unobstructed \SI{128}{nm} beam were
made before and after every TPB sample measurement, as described above. 
As stated in \cref{sec:data_taking}, the average
of the before- and after-contamination measurements was used to normalize
angular distributions taken at different
times, and the fractional difference between the two was incorporated as an
additional uncertainty on the corresponding angular distribution data.

\subsection{Degradation of TPB samples}
\label{sec:tpb_deg}
Although several tests prove the contamination light is a good measure of lamp
degradation even after many hours of ``lamp time'', this alone does not solve the problem of normalizing data taken on different samples at different times. If a sample is exposed to a significant amount of VUV light, its wavelength shifting efficiency will degrade and data taken before and after this degradation will no longer agree when normalized by contamination.

Tests show the degradation of TPB has a negligible effect on the angular
distribution of re-emitted light when the TPB has only been exposed to vacuum ultraviolet (VUV)
light during a few data taking periods. This was verified by taking data multiple times with
a previously unused sample and confirming the angular distributions were identical over
the course of these few measurements. To avoid effects caused by TPB
degradation, only samples that had minimal previous use were deployed for production data. Moreover, the production TPB samples were
stored in a dedicated vacuum chamber to avoid prolonged exposure to ambient
light and moisture. The samples used for production data had never been used
with the exception of three samples (SN104, SN112, SN113) listed
in~\cref{tab:list_of_tpb_samples} that were used once before.

\subsection{Indirect scattering of light}

With the lamp in operation, there can also be backgrounds from indirect
scattering of longer wavelength light off of reflective surfaces inside the
vacuum chamber. Both shifted light from the TPB and contamination light in the
beam can contribute to this background. This effect was estimated by shining
the \SI{128}{nm} beam onto an 0.635-\si{cm} thick
aluminum disk in the sample holder in the same configuration as
when taking TPB data. The magnitude of the effect was confirmed to be \(<
\SI{1}{\percent}\) of that seen with the PMT measuring TPB re-emission.

Because of the concern that reflective surfaces in the stainless steel vacuum
chamber may affect measurements, black Acktar ``Metal Velvet'' foil was used to
cover the inner walls of the chamber\cite{acktar}. The foil sheets are highly non-reflective
in the wavelengths of interest (\(\leq \SI{0.2}{\percent}\) specular
reflectance and \(\leq \SI{1}{\percent}\) hemispherical reflectance). No change
in the TPB re-emission measurement was observed before and after the foil
installation to within uncertainties.

\section{Results}
\label{sec:results}

\cref{tab:list_of_tpb_samples} summarizes the TPB film thicknesses used in this study. This range was chosen to span
the thickness at which previous works measured the greatest wavelength shifting
efficiency.

\begin{figure} [tbh!]
	\centering
	\includegraphics[width=1.1\linewidth, keepaspectratio]{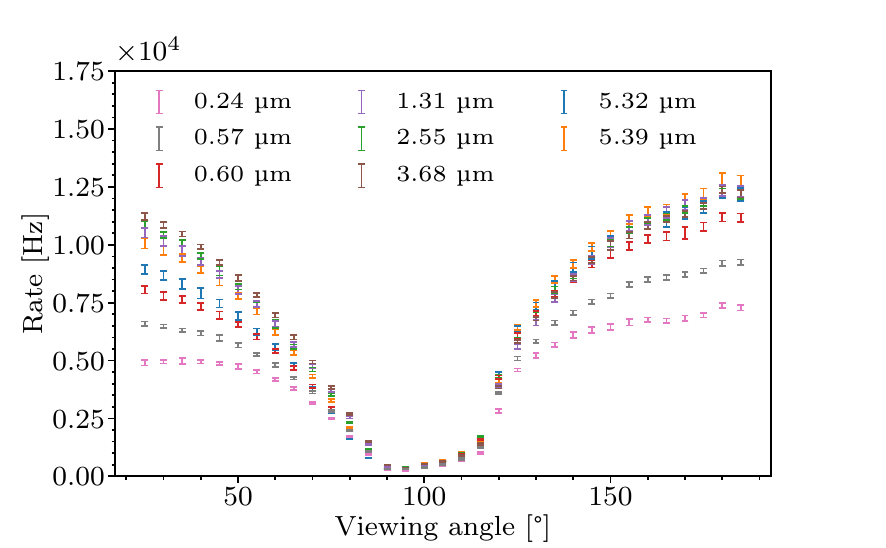}
	\caption{
		TPB re-emission intensity as a function of PMT viewing angle for
		a \SI{128}{nm} beam incident on TPB thin film samples of varying
		thickness.
		All the re-emission measurements are normalized to their corresponding contamination measurements.
	}
	\label{fig:tpb_reemission_intensity_vs_thickness}
\end{figure}

\cref{fig:tpb_reemission_intensity_vs_thickness} shows the angular
distributions measured from these samples when normalized to contamination as
discussed above. For PMT angles in the range \SIrange{90}{120}{\degree}, the line of sight from the TPB/acrylic sample to the PMT is obstructed by the back edge of the sample holder. This effect has been studied in MC and its impact on the measured distribution is well understood in terms of the thickness of the sample holder and angular alignment of the sample holder normal to the beam.

\begin{figure} [tbh!]
  \centering
  \includegraphics[width=1.1\linewidth, keepaspectratio]{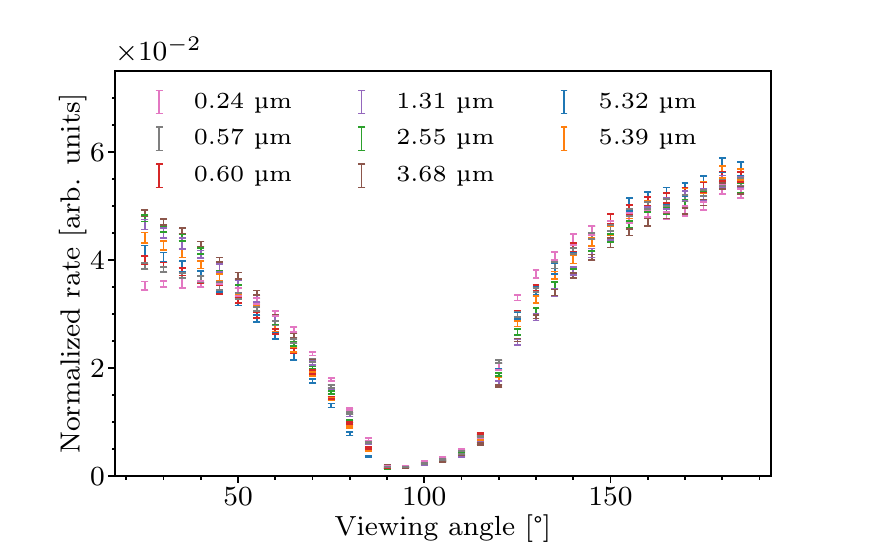}
  \caption{%
    Same data as in~\cref{fig:tpb_reemission_intensity_vs_thickness} but with
    each angular distribution normalized to unity.
  }%
  \label{fig:normalized_tpb_reemission_intensity_vs_thickness}
\end{figure}

An interesting pattern in these angular distributions is that more light is
detected \textit{behind} the TPB sample relative to the beam than on the same
side as the beam (described as transmitted and reflected light respectively). A similar asymmetry has been observed before when measuring re-emitted light
from TPB films on silicon substrates~\cite{Stolp:2016rde}. This asymmetry is much larger than can be accounted for by scattering of contamination light in the beam, which give photon rates of $\approx\SI{1}{kHz}$ exclusively in the narrow PMT viewing angles subtended by the beam. \cref{fig:normalized_tpb_reemission_intensity_vs_thickness}
shows the same data but with each angular distribution normalized to unity
in order to explicitly show the variation in shape. In theory, one expects this asymmetry to be less severe for thicker samples, in which the re-emitted light is more likely to undergo a second absorption if it proceeds towards the ``back" of the TPB sample, but this trend is not obviously apparent in the data.

\begin{figure} [t!]
	\centering
	\includegraphics[width=1.1\linewidth, keepaspectratio]{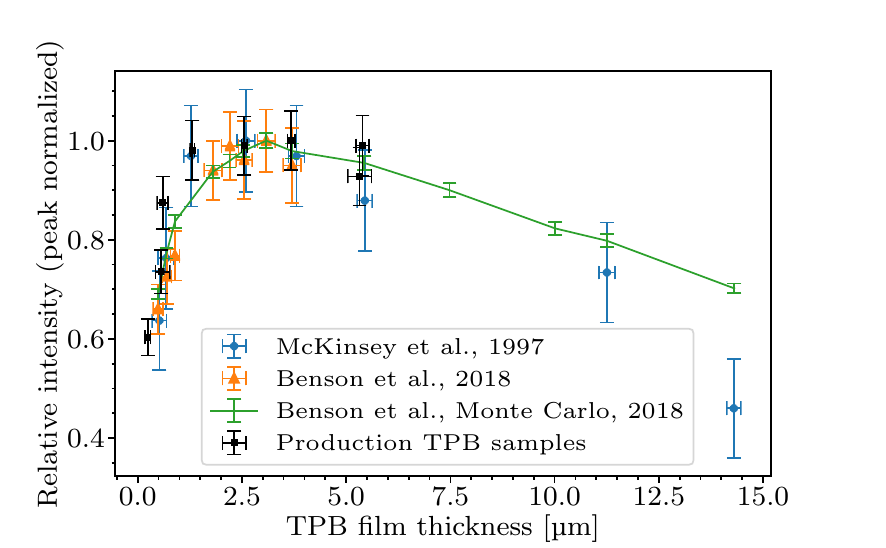}
	\caption{
		The integrated re-emission intensity of TPB thin film samples as a
		function of measured film thickness in this study is shown in black
		labeled ``Production TPB samples''. An overlay comparison is made with
		other previous studies. The peak values of all data sets shown are
		normalized to unity.
	}
	\label{fig:tpb_integrated_reemission_intensity_vs_thickness}
\end{figure}

\cref{fig:tpb_integrated_reemission_intensity_vs_thickness} shows the relative
integrals of these distributions as functions of sample thickness normalized to the peak.
Using $\chi^2$ instead of the peak changes the relative normalization of the data by about \SI{5}{\percent}.
The thickness corresponding to the peak wavelength shifting efficiency is
consistent with results of past measurements. Previous studies only looked at
the re-emitted light on one side of the TPB sample, so the observed agreement is
not necessarily trivial.

\section{Monte Carlo Studies}
\label{sec:monte_carlo}

Previous studies suggest the remission of light from the bulk of a TPB film happens isotropically~\cite{Benson:2017vbw,Stolp:2016rde}. However,
the angular distribution of re-emitted photons from a TPB film is theoretically affected by
the film thickness as mentioned above. Thicker films result in an angular
distribution more heavily populated at small angles, because fewer absorbed and
re-emitted photons are able to travel through the entire film without being
absorbed a second time. 

\begin{figure} [t]
	\centering
	\includegraphics[width=\linewidth, keepaspectratio]{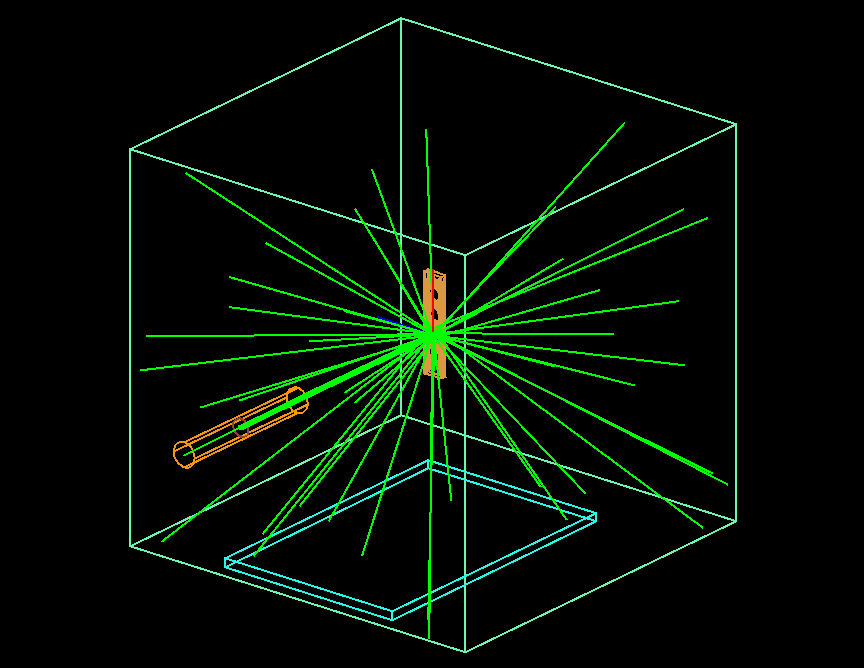}
	\caption{Rendering of the geometry used to simulate the angular distribution or re-emitted light from TPB films of various thicknesses. }
	\label{fig:sim3d}
\end{figure}

This sample-specific effect can in theory be leveraged to extract micro-physical properties of TPB by comparing angular distributions for multiple TPB film thicknesses to detailed MC simulations.

In this study, a micro-physical MC model of the experiment was made in an attempt to explore such an analysis. The simulation was built using the RAT Mini-CLEAN toolkit, which is a C++ wrapper for the popular GEANT4
particle transport simulation software. \cref{fig:sim3d} shows a
rendering of the model geometry. The optical model of the TPB used is based on
the model developed by Benson et al.~\cite{Benson:2017vbw} in connection with data they took in the same study. 

\begin{figure} [t]
	\centering
	\includegraphics[width=\linewidth, keepaspectratio]{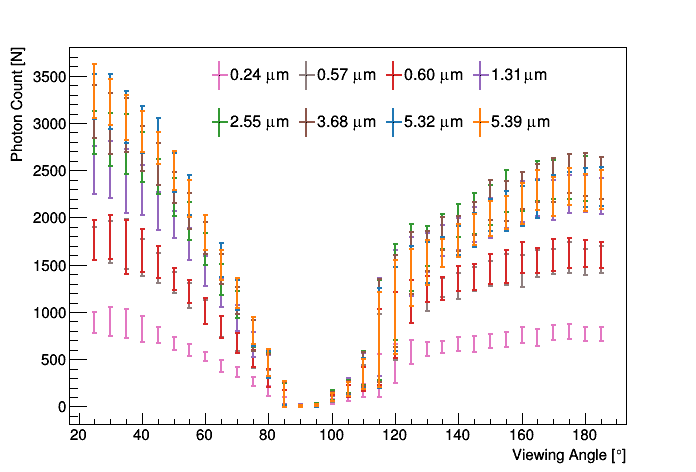}
	\caption{The angular distribution of re-emitted light from various TPB samples as simulated by a MC Model.}
	\label{fig:mc}
\end{figure}

\begin{figure} [hbt]
	\centering
	\includegraphics[width=\linewidth, keepaspectratio]{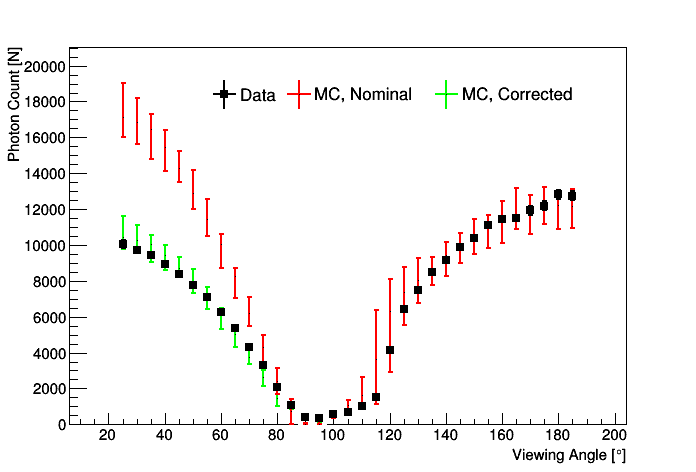}
	\caption{This figure illustrates the decent shape-agreement that can be induced between MC and data if the reflected part (left) of the MC angular distribution is scaled down. Red shows the nominal angular distribution produced by MC when normalized so the transmitted light in MC and Data agree. Green shows the decent shape agreement that can be achieved by scaling the reflected side of the angular distribution by some constant. This plot shows the angular distribution for a $\SI{5.39}{\micron}$ TPB sample. Note the choice to normalize to transmitted light and then ``correct" the integral of reflected light as opposed to the other way around is essentially arbitrary. The choice to normalize to transmitted light in this figure was made because most plausible physical explanations for the asymmetry involve additional attenuation of reflected light rather than increases in transmitted light.}
	\label{fig:corrected}
\end{figure}

The model geometry was validated by measuring the reflections of a $\SI{400}{nm}$ beam off a mirror in the sample holder
with the sample holder rotated to multiple beam-mirror incident angles, and comparing to simulated result. \cref{fig:mc} shows the MC prediction for the shape of the angular distribution of light emitted by the TPB for the same range of film thicknesses as presented in \cref{fig:tpb_reemission_intensity_vs_thickness}.  \cref{fig:corrected} shows a comparison of data to MC for the $\SI{5.39}{\micron}$ film thickness.

The error bars in \cref{fig:mc,,fig:corrected} include both statistical and systematic uncertainties having to do with the geometry of the apparatus. Systematic uncertainties were evaluated by measuring the effects of geometric parameters on the angular distributions in MC. The change in the angular distribution corresponding to a $1\sigma$ variation of a given geometrical parameter was added in quadrature to the total error on the MC angular distributions. In the apparatus, the intersection of the beam with a plane normal to the beam and containing the axis of the PMT rotation stage defined the origin. Among the geometric systematics accounted for, the most relevant were the displacement of the sample holder rotation axis relative to the origin, the displacement of sample holder from its own rotation axis, the angular displacement of the sample holder from normal to the beam, and lateral displacement of the axis of the PMT rotation stage relative to the origin.

With the MC geometry validated, it became clear the current optical model fails to reproduce some notable features of the measured angular
distributions. Namely, MC does not reproduce the aforementioned asymmetry between reflected and transmitted light in data. Furthermore, it was discovered that no plausible combination of adjustments to parameters in the optical model could reproduce the asymmetry.

Four parameters affecting the asymmetry between reflected and transmitted light in MC were thoroughly studied: the UV absorption length, the visible absorption length, the UV scattering length, and the visible scattering length. Here absorption lengths refer to the typical distance a photon travels through the TPB bulk before undergoing an inelastic interaction with a TPB molecule (e.g. shifting), and scattering lengths refer to the typical distance a photon travels through the TPB before undergoing an elastic interaction with a TPB molecule (i.e. no shifting). Naively one might suppose certain combinations of these parameters could produce a large asymmetry between reflected and transmitted light. For instance, one might think that a long UV absorption length, meaning the photons travel further into the TPB before being absorbed, combined with a short visible scattering length, meaning the re-emitted photons are likely to scatter in the TPB film before escaping, results in an attenuation of reflected light. This however is not the case; at large UV absorption lengths, it becomes equally likely for a UV photon to be absorbed at any depth into the TPB film and so no additional scattering of the re-emitted light can preference one direction over the other. This was confirmed in MC. It was observed that no combination of the mentioned parameters could produce an asymmetry of the magnitude seen in data, even when values several sigma away from the expected values were used.

One possible cause of the mis-modelling
could be poor handling of reflections and scattering at the surface of the TPB,
which scanning electron microscope (SEM) images show has a complicated topography. Another could be invalid use of the Rayleigh limit when
modelling scattering in the TPB, meaning it may be incorrect
to treat the scatterers as point particles.

Though it is unknown what optical processes lead to the asymmetry seen in data, it is possible to
induce decent agreement between MC and data by superficially
scaling the left side of the MC angular distribution as seen in \cref{fig:corrected}. This scaling is not the same when applying the correction to MC for different TPB thicknesses, nor does it appear to obey some simple function of TPB thickness.

One way to reproduce the attenuation of reflected light in MC is to construct a thin layer at the front of the TPB film which strongly scatters visible light. Although the choices of the thickness of the layer and the scattering length are arbitrary, it is found that a model like this can induce agreement between MC and data similar that in \cref{fig:corrected}.

\section{Conclusions}
\label{sec:conclusions}
We have presented measurements of the angular distribution of re-emitted light from TPB films
ranging in thickness from \SI{250}{nm} to \SI{5.5}{\micron} when exposed to $\SI{128}{nm}$ light in vacuum. Specifically, we measure photon rates at 32 PMT viewing angles from \SI{25}{\degree} to \SI{185}{\degree} as described in \cref{sec:apparatus}. This was done on 8 different samples as summarized in \cref{tab:list_of_tpb_samples}.

For all samples, we observed an asymmetry between the amount of re-emitted light that escaped the film in the backward (reflected) and the forward (transmitted) directions. We observed higher photon emission rates on the backside of the TPB film than on the front side. This is evident in \cref{fig:tpb_reemission_intensity_vs_thickness,fig:normalized_tpb_reemission_intensity_vs_thickness}.
The asymmetry is not
yet understood and presents the greatest challenge to performing a MC analysis with this data. The asymmetry could not be reproduced in MC by adjusting any of the bulk or surface optical properties available in the model. As described in \cref{sec:monte_carlo}, decent agreement between the shape of data and MC is observed for either the reflected or transmitted light individually, but the ratio of the integrals of the reflected and transmitted light do not match between the two. Good agreement between data and MC can be induced by applying a scaling to either the transmitted or reflected side of the angular distribution as seen in \cref{fig:corrected}. As compared to what is modeled in MC, it is unknown whether the asymmetry in data comes from additional attenuation of reflected light, reduced attenuation of transmitted light, or some combination of the two.

Some possible explanations for the asymmetry include surface effects which may cause additional scattering of visible light on the exposed surface of the TPB film, but not on the side attached to the acrylic substrate. This would further attenuate the reflected light. Another possibility may be that the effective size of the visible scatterers in TPB are on the order of crystal domains rather than TPB molecules, meaning a Rayleigh treatment of visible scattering is unjustified. Performing an analysis in the Mie scattering limit could increase the amount of transmitted light seen in MC.

The observed asymmetry in the angular distribution of re-emitted light may have implications for detector design. Broadly speaking the asymmetry can be taken advantage of in detector design by orienting films so more light is seen where desirable and less where undesirable. Future modeling of the asymmetry may lead to better micro-physical optical models of TPB for use in MC simulations of detectors that deploy TPB.
\section{Acknowledgements}
\label{sec:acknowledgements}

Construction and calibrations were supported by the Laboratory Directed Research and Development Program of Lawrence Berkeley National Lab under U.S. Department of Energy Contract No. DEAC02-05CH11231. Monte Carlo model development was supported by the U.S. Department of Energy, Office of Science, Office of High Energy Physics, under Award Number DE-SC0018974.

\printbibliography

\end{document}